\documentclass[%
reprint,
 superscriptaddress,
 amsmath,amssymb,
 aps,
 prl,
 floatfix,
]{revtex4-2}

\usepackage{graphicx}
\usepackage{dcolumn}
\usepackage{bm}
\usepackage{hyperref}

\usepackage[usenames,dvipsnames]{color}
\usepackage{colortbl}
\usepackage[english]{babel}
\newcommand{\REV}[1]{#1} 
\newcommand{\DEL}[1]{}

\begin{document}

\title{Strong-Field Theory of Attosecond Tunneling Microscopy}

\author{Boyang Ma}
\affiliation{Department of Physics, Technion---Israel Institute of Technology, Haifa 32000, Israel}
\affiliation{Solid State Institute, Technion---Israel Institute of Technology, Haifa 32000, Israel}
\affiliation{The Helen Diller Quantum Center, Technion---Israel Institute of Technology, Haifa 32000, Israel}

\author{Michael Kr\"uger}
\altaffiliation{Corresponding author: krueger@technion.ac.il}
\affiliation{Department of Physics, Technion---Israel Institute of Technology, Haifa 32000, Israel}
\affiliation{Solid State Institute, Technion---Israel Institute of Technology, Haifa 32000, Israel}
\affiliation{The Helen Diller Quantum Center, Technion---Israel Institute of Technology, Haifa 32000, Israel}

\date{\today}

\begin{abstract}
Attosecond observations of coherent electron dynamics in molecules and nanostructures can be achieved by combining conventional scanning tunneling microscopy (STM) with ultrashort femtosecond laser pulses. While experimental studies in the sub-cycle regime are underway, a robust strong-field theory description has remained elusive. Here we devise a model based on the strong-field approximation. Valid in all regimes, it provides a surprising analogy to the standard model of STM. We also show that the intuitive three-step model of attosecond science directly emerges from our model and describe the optimal conditions for attosecond STM experiments.
\end{abstract}

\maketitle



Scanning tunneling microscopy (STM) enables atomic-scale imaging of electronic states of surfaces and molecules~\cite{Binnig1982a, Binnig1982, Repp2005}, but lacks attosecond and femtosecond time resolution which characterizes electron dynamics in a wide range of systems, for example \REV{molecular charge migration}~\cite{Lepine2014, Calegari2016}. A straightforward approach to achieve time resolution in an STM is inducing and controlling tunneling currents between an atomically sharp metallic probe tip and a sample using phase-controlled ultrashort intense laser pulses. It is well known that the electric field waveform of such pulses can coherently control electron tunneling on the attosecond time scale in gases~\cite{Uiberacker2007, Shafir2012} and bulk solids~\cite{Vampa2015, You2017} and at metallic surfaces~\cite{Apolonski2004}, metallic nanotips~\cite{Kruger2011, Piglosiewicz2014, Kim2023, Dienstbier2023} and nanodevices~\cite{Rybka2016, Putnam2017, Ludwig2020, Bionta2021, Luo2023}. However, initial femtosecond STM experiments (see, e.g.,~\cite{Gerstner2000, Grafstrom2002, Merschdorf2002, Terada2010, Dey2013}) faced numerous experimental challenges and only recently, attosecond control of tunneling currents was achieved in an STM~\cite{Garg2020}. In the latter work, the authors were able to observe the transition from the multiphoton regime to the strong-field regime where tunneling occurs. Here, shifting the carrier-envelope phase (CEP) of 6\,fs near-infrared pulses assisted by a DC voltage enabled strong modulations of the tunneling current and the generation of attosecond bursts of tunneling electrons between the tip and a metallic sample. The basic mechanism behind the modulation is the strong sensitivity of tunneling to the instantaneous optical field, allowing for unipolar attosecond bursts of tunneling electrons inside an STM junction. This is very intriguing because it allows injecting or extracting electrons from a sample on the attosecond scale fully controlled by the laser field waveform alone, as shown already in the THz regime for the picosecond scale~\cite{Cocker2016}.

Existing strong-field theory models for ultrafast STM or, more generally, for ultrafast tunneling currents in nano-scale metal-insulator-metal (MIM) junctions are either analytical models, such as simple Fowler-Nordheim tunneling models~\cite{Rybka2016,Ludwig2020}, a strong-field model for the tip emission only without the barrier~\cite{Garg2020} or a quasi-classical imaginary time method calculation~\cite{Kim2021}, or {\em ab-initio} models, such as a numerical integration of the time-dependent Schr\"odinger equation (TDSE)~\cite{Garg2020} or time-dependent density functional theory (TDDFT) simulations~\cite{Ludwig2020,Hu2021,Bhan2022}. These models describe either partial aspects of the physics, such as the tunneling process, or provide exact results which are challenging to interpret. In this work, we introduce an intuitive and complete strong-field model for ultrafast laser-driven STM which provides deep insight into the underlying physical mechanisms. Applying the strong-field approximation (SFA)~\cite{Lewenstein1994, Milosevic2006, Amini2019} to our problem, we find the intriguing three-step picture~\cite{Corkum1993,Kulander1993} which beautifully describes the hallmark recollision effects of attosecond science, high-harmonic generation (HHG,~\cite{Ferray1988, Vampa2015}), high-order above-threshold ionization (ATI,~\cite{Paulus1994a}) and photoemission (ATP,~\cite{Kruger2011}), as well as non-sequential double ionization~\cite{Fittinghoff1992}. We identify an energy cutoff law, which is markedly different from the former effects, and describe the optimal conditions for performing attosecond tunneling experiments in an STM.

\begin{figure}
\includegraphics[width=1\columnwidth]{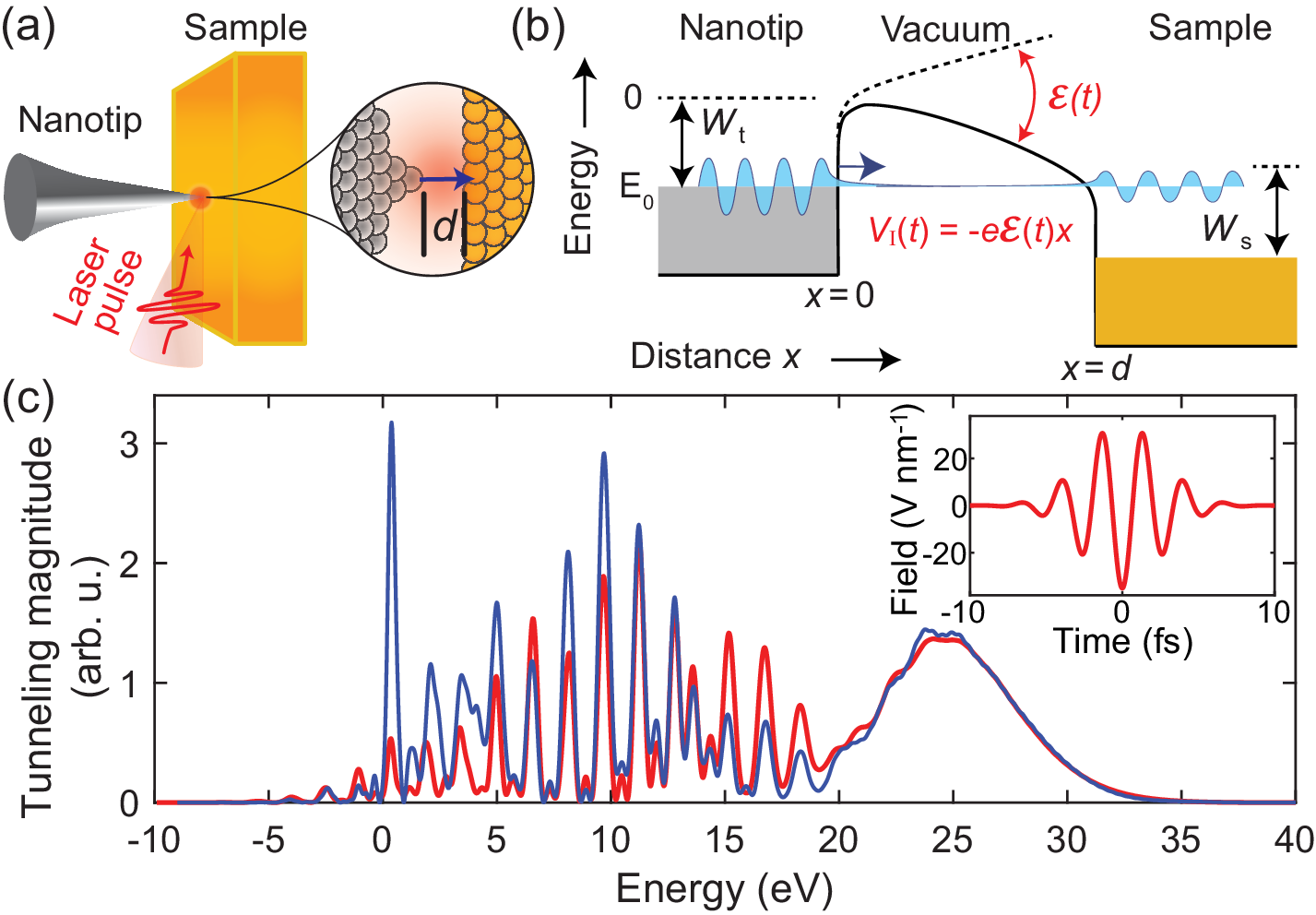}
\caption{\label{fig:epsart} Strong-field theory for ultrafast STM. (a) A schematic of the ultrafast STM. (b) One-dimensional potential model of the laser-driven STM junction\REV{ ($W_\mathrm{t}$: tip work function, $W_\mathrm{s}$: sample work function, ${\cal E}$: laser electric field)}. (c) The tunneling electron spectrum calculated by the TDSE (red) and our SFA theory (blue) following Eq.~\ref{eq2}.  Here, $d = 1$\,nm, the pulse duration of the laser is 6\,fs (full width at half maximum\REV{, see inset for the field waveform}) and the field amplitude $F$ is $35\,\mathrm{V\,nm}^{-1}$. The material for tip and sample is gold \REV{($W_\mathrm{t} = W_\mathrm{s} = 5$\,eV), the tip electron is initially at the Fermi energy $E_0 = -W_\mathrm{t} = -5$\,eV.} Static fields are absent.}
\label{fig1}
\end{figure}

\REV{The ultrafast STM can be described as a static MIM junction illuminated by a near-infrared (NIR) femtosecond laser pulse (see Fig.~\ref{fig1}(a) and (b)). We assume that the laser field is screened perfectly inside the tip and sample and is only present in the vacuum gap. When the laser turns on, the time-varying laser electric field drives electrons inside the junction, generating a laser-induced tunneling (LIT) current. Even though the STM is limited by the electronic response time, the time-integrated LIT current can be measured and recorded by the STM. Under the laser interaction, currents inside the STM can flow from the tip to the sample and vice versa. They can be thought of as independent current flows and can be modeled in the same way but with the opposite sign of the electric field (see Supplemental Material~\cite{NoteSuppl}\nocite{Yoshioka2016,Jelic2017,Luo2021,Smirnova2014,Pedatzur2015,Ott2013,Hommelhoff2006}). Therefore, for simplicity, we only focus on the current flow from the tip to the sample. Furthermore, the extremely sharp nanotip and the sample can generate a near-field in the gap, localizing the LIT current only at the tip apex~\cite{Garg2020}. Therefore, despite its simplicity, a one-dimensional treatment promises accurate calculations~\cite{Yoshioka2016,Jelic2017,Garg2020}. We define a spatial coordinate $x$ and a junction width $d$. The tip-vacuum boundary is located at $x=0$ (see Fig.~\ref{fig1}(b)). In this letter, we study the time-integrated LIT current by analyzing the tunneling amplitude after the interaction (see Supplemental Material for detailed derivations~\cite{NoteSuppl}).} 

\REV{In the conventional STM, the tunneling amplitude\DEL{the tip and sample form a static MIM junction which} can be described using Bardeen's tunneling theory~\cite{Bardeen1961}, the standard theory approach for STM~\cite{Chen2021}. \DEL{In the one-dimensional case, we define a spatial coordinate $x$ and a junction width $d$. The tip-vacuum boundary is located at $x=0$.}By defining the eigenfunction of the tip as $\Psi_{\mathrm{t}}$ and the eigenfunction of the sample at energy state $E$ as $\psi_{E}$, the static tunneling current from tip to sample is proportional to the absolute square of Bardeen's tunneling matrix element $C_\mathrm{stat}$, which is given by}
\begin{equation}
\begin{split}
 C_\mathrm{stat}=\frac{\hbar^{2}}{2m} \bigg[\Psi_{\mathrm{t}}(x)\frac{\partial}{\partial x}\psi^{*}_{{E}}(x)-\psi^{*}_{{E}}(x)\frac{\partial}{\partial x}\Psi_{\mathrm{t}}(x)\bigg]_{x=d}.
 \label{eqA1}
\end{split}
\end{equation}
\REV{Here $\hbar$ is the reduced Planck constant and $m$ the electron mass. \DEL{$\Psi_{\mathrm{t}}$ is an eigenfunction of the tip and $\psi^{*}_{{E}}$ is the conjugate of an eigenfunction of the sample, respectively.}The superscript star represents the complex conjugate. In Bardeen's theory, the wave functions are assumed to be orthogonal to each other. The term in the brackets is evaluated at the boundary $x=d$, hence only the information of the wave functions at the vacuum-sample boundary is required while explicit information about the barrier potential\DEL{and static voltage} remains absent. Despite its simplicity, Bardeen's tunneling theory enables quantitative modeling of STM (see Ref.~\cite{Chen2021} for details).}

\REV{In the ultrafast STM, we need to solve a time-dependent problem involving a strong laser field to obtain the LIT current. Here we formulate the LIT amplitude $M_E$ in a form similar to $C_\mathrm{stat}$ (Eq.~\ref{eqA1}), which is given by}
\begin{equation}
\begin{split}
M_{E}= \frac{i\hbar}{2m}\int_{-\infty}^{\infty} \bigg[\Psi_{\mathrm{Is}}(x,t)\frac{\partial}{\partial x}\psi^{*}(x,t) \\
-\,\psi^{*}(x,t)\frac{\partial}{\partial x}\Psi_{\mathrm{Is}}(x,t)\bigg]\bigg\vert^{x=d}_{x=0}\;dt.
 \label{eq1}
\end{split}
\end{equation}
\REV{We have obtained $M_E$ from expansions of the Dyson equation of the time-dependent Schr\"odinger equation~\cite{Milosevic2006, Ivanov2005} (see the Supplemental Material for the complete derivation~\cite{NoteSuppl}). $\Psi_{\mathrm{Is}}(x,t)$ is the time-propagated wave function driven by the laser, which is generated from $\left| \Psi_0 \right\rangle$ inside the gap, where $\left| \Psi_{0} \right\rangle$ is the initial state of the system with an electron at the Fermi energy $E_0$ of the tip. The subscript ``$\mathrm{Is}$'' represents the interaction with the static and time-dependent potentials  (see Supplemental Material~\cite{NoteSuppl}). $\psi^{*}(x,t)=\left\langle \psi_E \right| \REV{\chi_\mathrm{s}} \; U(\infty,t) \left| x \right\rangle$ is nothing else than the complex conjugate of the wave function in the sample region propagated to a specific time. Here, we denote the full time evolution operator of the system as $U(t_2,t_1)$, with $t_2 > t_1$, and define $\chi_\mathrm{s}$ to be 1 for $x \geq d$ and 0 elsewhere, which is an operator to select only the wave function in the sample region. The notation $[...]\big\vert^{x=d}_{x=0}$ at the end of Eq.~\ref{eq1} stands for the subtraction of the term inside the brackets evaluated at $x=0$ from the term evaluated at $x=d$.}

\REV{The main difference to the Bardeen's expression is the fact that, in $M_E$, the matrix element is now time-dependent, requiring a time integration. We also note that Eq.~\ref{eq1} also includes the tip-vacuum boundary at $x=0$. $\Psi_{\mathrm{Is}}(x,t)$, which is driven back and forth in the laser field, can scatter at the tip boundary at $x=0$, leading to a rescattering effect~\cite{Kruger2011}. The LIT amplitude depends on the total flux through the junction boundaries, which is a direct result of the flux continuity (see Supplemental Material~\cite{NoteSuppl}). We find excellent agreement between solutions of Eq.~\ref{eq1} and a numerical solution of the full TDSE using the Crank-Nicolson scheme \cite{Yalunin2011,Kruger2011}~(see Supplemental Material~\cite{NoteSuppl}).}

The presence of the time-dependent wave function\REV{s} $\psi^{*}(x,t)$ \REV{and $\Psi_{\mathrm{Is}}(x,t)$} in Eq.~\ref{eq1} \DEL{is} \REV{are} hindering us from studying the tunneling dynamics in a strong ultrashort laser field analytically. We therefore apply the SFA\REV{, which treats the \REV{laser-driven (continuum) }electron \DEL{in the continuum }as a free particle moving in the laser field (Volkov wave), neglecting any influence of the field-free potential~\cite{Lewenstein1994, Milosevic2006, Amini2019}. In our specific case, we use two assumptions to \DEL{fulfill}\REV{apply} the SFA. First, we neglect the effect of the image charge potential in the vacuum junction. The image charge is induced when an electron is present inside the junction, causing an attractive force towards the surfaces~\cite{Pitarke1990}. However, for laser field strengths on the order $10\,\mathrm{V\,nm}^{-1}$ and higher, this force is negligible compared to the force exerted by the laser field. Second, we ignore reflections of the wavefunction from the tip boundary $x=0$ where the field-free potential has a high step. Such reflections have been observed at nanotips and are known as rescattering. Here an electron is emitted from the tip, but returns to the tip surface after about $3/4$ of an optical cycle and undergoes elastic scattering~\cite{Kruger2011,Wachter2012}. However, the contribution from rescattering to the total current in those experiments is weak. Moreover, in strong fields on the order $10\,\mathrm{V\,nm}^{-1}$ and higher, most electrons cross the tiny STM junction in less than an optical half-cycle, hence the electron does not quiver and rescattering cannot take place~\cite{Ludwig2020}. In addition to these two assumptions, we also treat the sample region as an isolated system so that the wave function component which has been transmitted to the sample region \REV{evolves }freely. This is justified by the fact that the electric field is strongly screened and hence there is no time-dependent force present in the sample region. }\DEL{The SFA fully disregards the effect of the binding potential outside the metals. }Therefore, we replace $\psi^{*}(x,t)$ \REV{and $\Psi_{\mathrm{Is}}(x,t)$} in Eq.~\ref{eq1} with the eigenfunction $\psi^{*}_{E}(x,t)$ \REV{and a superposition of Volkov wavefunctions $\Psi_{\mathrm{V}}(x,t)$, respectively,} and obtain the SFA expression 

\begin{equation}
\begin{split}
M_{E,\mathrm{SFA}}=\frac{i\hbar}{2m}\int_{-\infty}^{\infty} \bigg[\Psi_\mathrm{V}(x,t)\frac{\partial}{\partial x}\psi^{*}_{E}(x,t) \\
-\,\psi^{*}_{E}(x,t)\frac{\partial}{\partial x}\Psi_\mathrm{V}(x,t) \bigg]_{x=d}\;dt\,.
 \label{eq2}
\end{split}
\end{equation}

The laser field inside the junction excites the initial wave function from its ground state to \REV{a }laser-driven Volkov continuum state and moves the excited wave function to the sample region, where this transport finishes by projection on the energy eigenstates of the sample. Figure~\ref{fig1}\REV{(c)} shows a comparison between Eq.~\ref{eq2} and a numerical integration \REV{of the TDSE (see Supplemental Material~\cite{NoteSuppl} for details of the TDSE)}. We assume a peak field of $35\,\mathrm{V\,nm}^{-1}$ for a 6-fs laser pulse at a central wavelength of 830\,nm. The width of the STM junction between a gold nanotip and a gold sample is chosen as $d = 1$\,nm. We obtain excellent agreement between our strong-field model and the TDSE result.

\begin{figure}
\includegraphics[width=1\columnwidth]{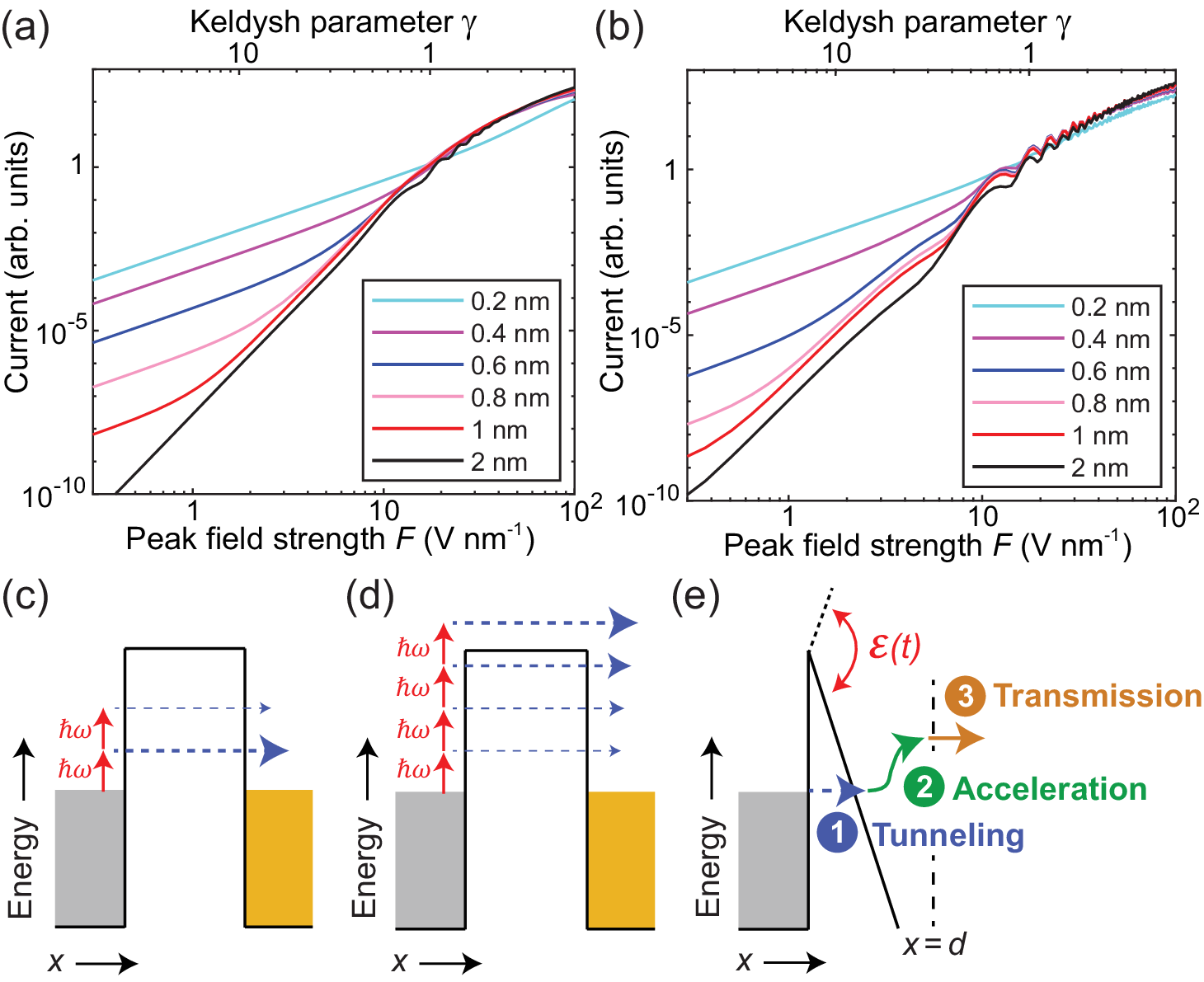}
\caption{Regimes of ultrafast STM. (a) Double-logarithmic plot of LIT current vs.~light peak field strength $F$ and Keldysh parameter $\gamma$ for different barrier widths $d$. The calculation is based on a numerical integration of the time-dependent Schr\"odinger equation for 6-fs pulses at 830\,nm wavelength and a gold nanotip and sample. (b) The same plot calculated with the strong-field theory model (Eq.~\ref{eq2}). (c) Weak-field regime with dominating one-photon excitation and subsequent static tunneling. (d) Multiphoton regime with subsequent static tunneling or over-barrier emission. (e) Field-driven tunneling regime with three steps -- laser-driven tunneling, acceleration and transmission to the sample region.}
\label{fig2}
\end{figure}

Figure~\ref{fig2} shows the total LIT current vs field strength, revealing the different regimes of ultrafast STM depending on $F$ and $d$. We also show the Keldysh parameter $\gamma=({\omega\sqrt{2m|E_{0}|}}) / ({|e|F})$~\cite{Keldysh1965}. TDSE calculation (Fig.~\ref{fig2}(a)) and our strong-field model (Fig.~\ref{fig2}(b)) are in good agreement. At $\gamma \gg 1$, the current scales linearly with intensity for $d \leq 1\,$nm. Here, a single photon is absorbed and static tunneling enables a transmission of the thus excited electron to the sample region (Fig.~\ref{fig2}(c)). At increasing field strengths, but still at $\gamma > 1$, a transition to multiphoton excitation take place. Electrons that have absorbed more than one photon are either transmitted through static tunneling or over-barrier propagation (Fig.~\ref{fig2}(d)). Notably, for a very thin barrier of 0.2\,nm this transition to multiphoton excitation does not take place because the static tunneling remains extremely efficient. These phenomena have already been studied theoretically and experimentally based on weak perturbation methods~\cite{Garg2021, Garg2022}. At $\gamma \sim 1$, we find the transition to the tunneling regime, in great similarity to nanotip photoemission, for instance~\cite{Bormann2010,Dombi2010,Kruger2011,Yalunin2011,Kruger2018,Dombi2020,Dienstbier2023}. The curves for all barrier widths start to merge here because the laser field completely dominates the process and static tunneling does not have much of a role anymore for $\gamma < 1$. We note that our strong-field model magnifies the channel-closing effect with its strong oscillations around $\gamma=1$, which should be only strong for large barrier widths according to the TDSE results.

Writing Eq.~\ref{eq2} more explicitly, we find a more intuitive SFA expression for the tunneling amplitude,

\begin{equation}
\begin{split}
M_{E,\mathrm{SFA}}=\int_{-\infty}^{\infty}\int_{-\infty}^{t_{2}}\sqrt{\frac{i}{8\pi m\hbar^3(t_{2}-t_{1})}} 
\eta(t_{2})\xi(t_{1}) \\ 
 \times \exp \left( {\frac{i}{\hbar}S(t_{2},t_{1})} \right) \;dt_{1}dt_{2},
 \label{eq3}
 \end{split}
\end{equation}
with the semi-classical action 
\begin{equation}
\begin{split}
S(t_{2},t_{1}) &= E t_{2}+\frac{{\tilde p}^2}{2m}(t_{2}-t_{1}) \\ &-\int_{t_{1}}^{t{2}}\frac{e^2A^2(\tau)}{2m} \;d\tau\,+\vert E_{0}\vert t_{1},
\end{split}
 \label{eq4}
\end{equation}
where $e$ is the (negative) electron charge, $\tilde p = ({\int_{t_{1}}^{t_{2}}{eA(\tau)}\;d\tau\,+md}) / ({t_{2}-t_{1}})$ is the canonical momentum and $A(t)$ is the vector potential of the light field which has a relation to the electric field ${\cal E}(t)=-{\partial A(t)} / {\partial t}$. The prefactors $\xi(t_{1})$ and $\eta(t_{2})$ describe normalization factors and transition matrix elements at times $t_1$  and $t_2$ (see~\REV{Supplemental Material}~\cite{NoteSuppl}). The time integrals in Eq.~\ref{eq3} and~\ref{eq4} can be analyzed using the saddle point method~\cite{Lewenstein1994,Salieres2001,Ivanov2005,Yalunin2011}. We obtain the saddle point equations

\begin{eqnarray}
\frac{\left[\tilde p_\mathrm{s}-eA(t_\mathrm{1s})\right]^2}{2m}=-\left\vert E_0\right\vert,
\\
\int_{t_\mathrm{1s}}^{t_\mathrm{2s}}\frac{\left[\tilde p_\mathrm{s}-eA(\tau)\right]}{m}\;d\tau\,=d,
\\
\frac{\left[\tilde p_\mathrm{s}-eA(t_\mathrm{2s})\right]^2}{2m}=E,
\end{eqnarray}
where $\tilde p_\mathrm{s}$ is an effective canonical momentum which is conserved in the homogeneous electric field, and $t_\mathrm{1s}$ and $t_\mathrm{2s}$ are the saddle points of emission time and arrival time, respectively.

The above three saddle point equations give us a direct semi-classical scenario of the electron transport in ultrafast STM within the three-step trajectory picture, illustrated in Fig.~\ref{fig2}(e): (1) emission by tunneling into the vacuum gap (Eq.~\REV{6}), (2) acceleration by laser field across the gap (Eq.~\REV{7}), and (3) transmission into the sample region (Eq.~\REV{8}). The first saddle point equation enforces energy conservation at the emission time $t_{1s}$. The initial energy is negative, which leads to complex values for $t_\mathrm{1s}$, $t_\mathrm{2s}$ and $\tilde p_{s}$. The imaginary part of the emission time is due to the classically forbidden dynamics under the barrier and can be interpreted as tunneling time~\cite{Lewenstein1994,Zheltikov2016}. The second equation describes the displacement of the quasi-free electron accelerated by a homogeneous electric field. The electron starts moving at the emission time $t_\mathrm{1s}$, but unlike the hallmark three-step recollision processes of attosecond science, we find in Eq.~\REV{7} that the electron does not return to the boundary $x = 0$, hence there is no recollision with the parent matter. Instead, it moves to the second boundary $x = d$. The last saddle point equation represents the energy conservation at the final time $t_\mathrm{2s}$ as the electron is transmitted into the sample region with a final energy $E$.

\begin{figure}
\includegraphics[width=1\columnwidth]{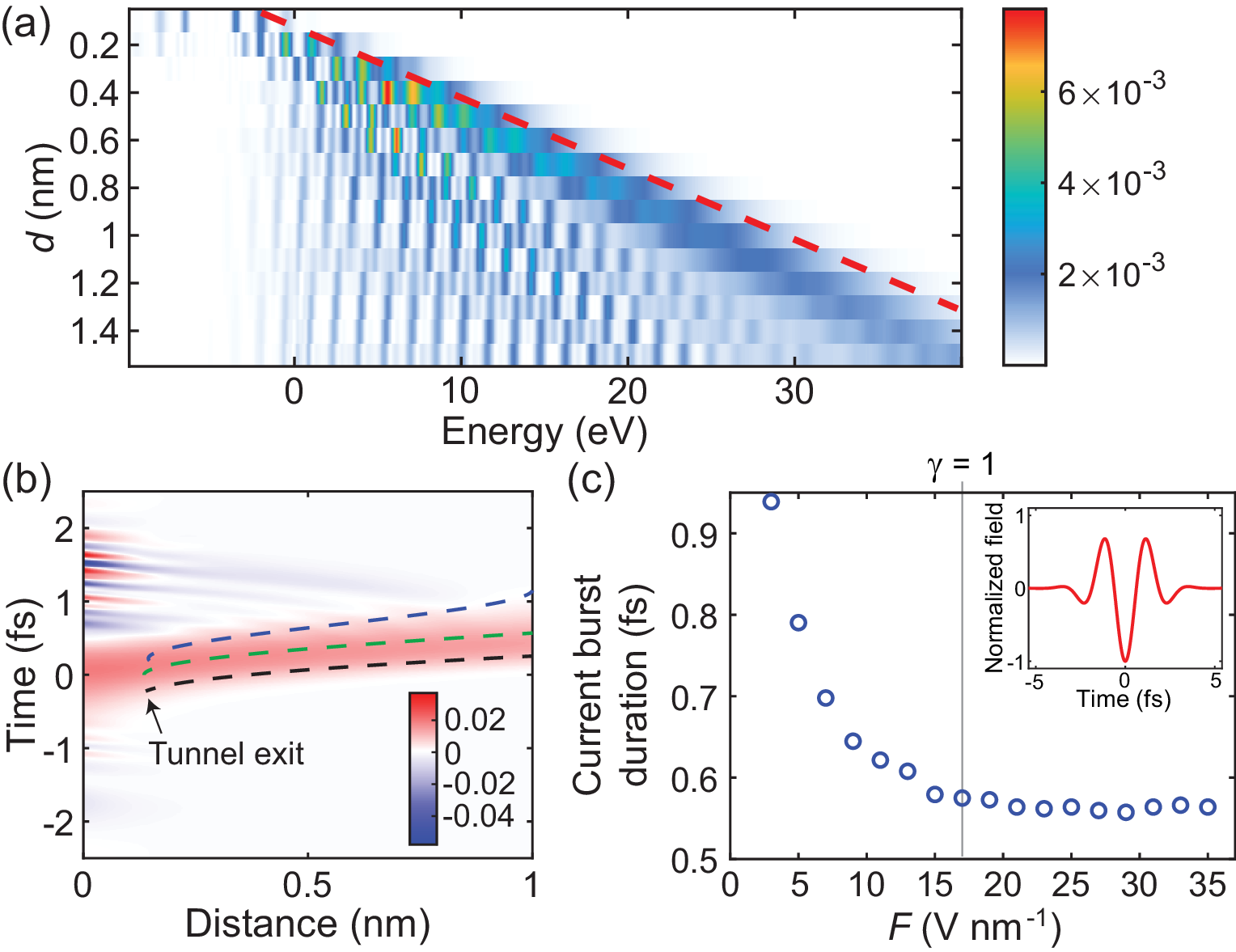}
\caption{Cutoff and attosecond transport. (a) Spectrum of the tunneling electron as a function of $d$ for the same parameters as Fig.~\ref{fig1}(c) calculated with the TDSE. The red curve indicates the cutoff law derived from the saddle point equations. (b) Color plot of local currents as a function of space and time for $d = 1\,$nm, $F = 35\,\mathrm{V\,nm}^{-1}$ and a 3-fs single-cycle laser pulse. From top to bottom, three curves show selected electron trajectories derived from the saddle-point equations for final energies of $\sim 0$\,eV, 15\,eV and 30\,eV (cutoff), respectively. The laser field peaks at $t = 0$\,fs. (c) Duration of the current burst as a function of $F$.\REV{ Inset: Field wave form in (b) and (c).}}
\label{fig3}
\end{figure}

\begin{figure}
\includegraphics[width=0.9\columnwidth]{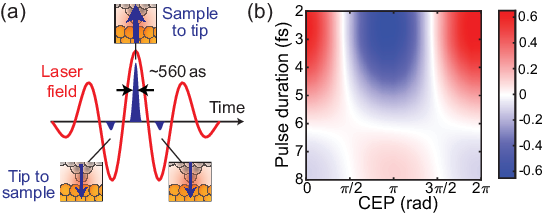}
\caption{Symmetry breaking of tunneling solely by the laser field waveform. (a) Illustration of current rectification by a sub-two-cycle laser pulse. (b) Dependence of the total tunneling current and its sign on the carrier-envelope phase (CEP) for different pulse durations ($F = 17\,\mathrm{V\,nm}^{-1}$, $\gamma \sim 1$ and $d = 1$\,nm). Strongly unipolar current bursts are found for 4\,fs duration and shorter.}
\label{fig4}
\end{figure}

The saddle point equations can also be used to derive the classical cutoff energy. For a continuous wave light field, we obtain a law for cutoff energy of the form $E_\mathrm{cutoff} = |e|(F_{0}+F)d + E_0$, where $F_0$ is the static field that is usually present in an STM (see Supplemental Material~\cite{NoteSuppl}). Indeed, the TDSE simulations in Fig.~\ref{fig3}(a) show that the tunneling spectrum beyond the classical cutoff is strongly attenuated. Our result stands in contrast to the cutoff laws of gas-phase HHG, ATI and ATP which depend on the ponderomotive energy which in turn is dependent on intensity and frequency. For ultrafast STM, the \REV{decisive quantities are} the total electric field $F_{0}+F$ and the barrier width $d$, but there is no dependence on the frequency $\omega$. The thin junction of the STM ensures that the transport finishes in a short time without any quiver motion, in analogy to near-field acceleration at nanotips~\cite{Herink2012,Ciappina2013} and nanojunctions~\cite{Ludwig2020}. Figure~\ref{fig3}(b) shows the local current density as a function of space and time inside a 3-fs light pulse at $35\,\mathrm{V\,nm}^{-1}$ ($\gamma \sim 0.5$). Indeed, a single burst is generated at the central field crest at $t = 0$ and travels to the sample on the attosecond time scale. The dynamics of the burst correspond well with electron trajectories calculated using the saddle point equations (see Supplemental Material~\cite{NoteSuppl} for calculation details). Attosecond currents can be transported to the sample region within less than a femtosecond. The duration of the laser-driven tunneling burst is strongly connected to the Keldysh parameter (see Fig.~\ref{fig3}(c)). As we increase the field strength and move from the multiphoton regime to the non-adiabatic tunneling regime~\cite{Yudin2001}, the duration decreases until it saturates at $\gamma \sim 1$ to a value of $\sim 560$\,as where tunneling occurs adiabatically. This suggests that attosecond experiments can also be carried out in the non-adiabatic tunneling regime ($1 \lesssim \gamma \lesssim 4$) where the energy spread of the tunneling electron wavepacket is more limited. A sufficiently short CEP-stable laser can break the symmetry between the two directions of current flow between tip and sample, inducing current rectification controlled by the CEP (see Fig.~\ref{fig4}(a) for an illustration). For instance, this ability may enable the extraction of an electron from a molecular sample driven by a CEP-stable pump pulse, followed by the injection of an electron into the remaining hole by a probe pulse with inverted CEP, opening up spatio-temporal imaging of ultrafast charge migration~\cite{Calegari2016}. Such an experiment is challenging because it requires near-single-cycle pulses in order to achieve strong symmetry breaking solely by the laser field in the absence of a static field (see Fig.~\ref{fig4}(b) and Supplemental Material~\cite{NoteSuppl}). However, experimental progress in single-cycle laser development and CEP-controlled currents in nano-junctions~\cite{Rybka2016,Ludwig2020} suggest that such symmetry breaking can be implemented also in an ultrafast STM.

In conclusion, we have introduced a strong-field theory for attosecond electron transport in ultrafast STM and other MIM junctions. Obtaining good agreement with an exact solution of the TDSE, we shed light on the transport mechanisms and the underlying temporal and spectral dynamics. We find the three-step picture at the heart of the tunneling regime of ultrafast STM. Possessing a strong versatility like other strong-field theories in attosecond science, our theory can be extended to include different materials, the density of states of sample and tip, the static field, image charge potential effects and the optical near-field enhancement inside the vacuum gap. We expect that our work will open the door to a rich interplay between theory and experiment at the extreme frontiers of spatio-temporal microscopy. \REV{Ultrafast pump-probe STM with phase-controlled current bursts will enable triggering and probing ultrafast charge dynamics in molecules, nanostructures and defect states, where a pump-probe delay-dependent, spatially localized current signal reveals charge oscillations on the femtosecond and attosecond time scales.}

The authors acknowledge Zhaopin Chen and Yiming Pan for insightful discussions. This project has received funding from the European Union's Horizon 2020 research and innovation program under grant agreement No 853393-ERC-ATTIDA and from the Israel Science Foundation under grant agreement No 1504/20. We also acknowledge the Helen Diller Quantum Center at the Technion for partial financial support.


%

\end{document}


\title{Strong-Field Theory of Attosecond Tunneling Microscopy \\
-- Supplemental Material --}

\author{Boyang Ma}
\affiliation{Department of Physics, Technion---Israel Institute of Technology, Haifa 32000, Israel}
\affiliation{Solid State Institute, Technion---Israel Institute of Technology, Haifa 32000, Israel}
\affiliation{The Helen Diller Quantum Center, Technion---Israel Institute of Technology, Haifa 32000, Israel}

\author{Michael Kr\"uger}
\altaffiliation{Corresponding author: krueger@technion.ac.il}
\affiliation{Department of Physics, Technion---Israel Institute of Technology, Haifa 32000, Israel}
\affiliation{Solid State Institute, Technion---Israel Institute of Technology, Haifa 32000, Israel}
\affiliation{The Helen Diller Quantum Center, Technion---Israel Institute of Technology, Haifa 32000, Israel}

\date{\today}

\maketitle

\section{Induced currents in the laser-driven scanning tunneling microscope}
\label{sec:1}

A conventional scanning tunneling microscope (STM) consists of a conductive nanotip and a sample separated by a vacuum gap. Within a simple description of the STM using the Schr\"odinger equation, the metal-vacuum-metal sandwich structure can be described as two potential wells separated by a thin vacuum potential barrier. Therefore, wavefunctions located in the tip and sample can bifurcate into a tunneling wave and a reflected wave separately in these two potential wells after evolution. The continuity equation $({\partial} / {\partial t})\vert\Psi\vert^2+\nabla\cdot J=0$ should be satisfied by the evolving wavefunctions in any region and at any time. Integrating it over time and space, we can derive the time-accumulated currents on the boundaries of a specific region and evaluate the probability difference between the final time and initial time in this region:

\begin{equation}
\begin{split}\label{eq1}
\int_{-\infty}^{+\infty}J_{2}\;dt\,-\int_{-\infty}^{+\infty}J_{1}\;dt\,=-\bigg[\left\langle \Psi (+\infty) \right\vert \;\chi_{(x_{1}, x_{2})}\; \left\vert  \Psi (+\infty) \right\rangle-\left\langle \Psi (-\infty) \right\vert \;\chi_{(x_{1},x_{2})}\; \left\vert  \Psi (-\infty) \right\rangle\bigg].
\end{split}
\end{equation}

Here we have used $-\infty$ and $+\infty$ to denote the initial time and the final time. We define an operator $\chi_{(x_{1}, x_{2})}$ in order to select values in $x \in (x_{1}, x_{2})$. Inside the vacuum gap, the  wavefunctions will return to their initial state after a sufficiently long time. Therefore, we obtain the flux conservation of currents ($J_\mathrm{t}$ and $J_\mathrm{s}$) from the tip and the sample, respectively:

\begin{equation}
\begin{split}\label{eq2}
\int_{-\infty}^{+\infty}J_{s}\;dt\,=\int_{-\infty}^{+\infty}J_{t}\;dt\,.
\end{split}
\end{equation}

To obtain the total current, we apply Eq.~\ref{eq1} to the sample region and let $ J_{\infty}=0$. The wavefunction in the tip (or sample) $\vert\Psi_{t}\rangle$ ($\vert\Psi_{s}\rangle$) bifurcates into a tunneling wavefunction $\vert\Psi_{tT}\rangle$ ($\vert\Psi_{sT}\rangle$) in the sample (tip) and a reflected wavefunction $\vert\Psi_{tR}\rangle$ ($\vert\Psi_{sR}\rangle$) in the tip (sample). Since the electrons emitted from the tip and the sample can have a random global phase, we multiply a phase factor $\exp{(i\phi)}$ to the $\vert\Psi_{s}\rangle$. Considering $N$ laser pulses, the average current per pulse is 

\begin{eqnarray}\label{eq3}
{\cal J}&=&\frac{1}{N}\sum_{n=1}^{N}\int_{-\infty}^{+\infty}J_{s}\;dt\,\nonumber\\
&=&\frac{1}{N}\sum_{n=1}^{N}\{\langle \Psi_{tT}+\Psi_{sR}e^{i\phi(n)}\vert \Psi_{tT}+\Psi_{sR}e^{i\phi(n)}\rangle\ -\langle \Psi_{s} \vert  \Psi_{s} \rangle\}\nonumber\\
&=&\frac{1}{N}\sum_{n=1}^{N}\{\langle \Psi_{tT}+\Psi_{sR}e^{i\phi(n)} \vert\Psi_{tT}+\Psi_{sR}e^{i\phi(n)}\rangle\\
&&-\langle \Psi_{sT}+\Psi_{sR} \vert e^{-i\phi(n)} U(+\infty,-\infty)U^{\dagger}(+\infty,-\infty)e^{i\phi(n)}\vert\Psi_{sT}+\Psi_{sR}\rangle\}\nonumber\\
&=&\frac{1}{N}\sum_{n=1}^{N}\{\ \langle \Psi_{tT} \vert\Psi_{tT}\rangle-\langle \Psi_{sT} \vert\Psi_{sT}\rangle+2\Re(\ e^{i\phi(n)}\langle \Psi_{tT} \vert\Psi_{sR}\rangle\ )\ \}\nonumber\\
&=&\langle \Psi_{tT} \vert\Psi_{tT}\rangle-\langle \Psi_{sT} \vert\Psi_{sT}\rangle\nonumber
\end{eqnarray}

Equation~\ref{eq3} shows that the contributions to the current are from the tunneling probabilities in the tip and the sample. The other contributions are averaged out.

\begin{figure*}
\includegraphics[width=0.5\textwidth]{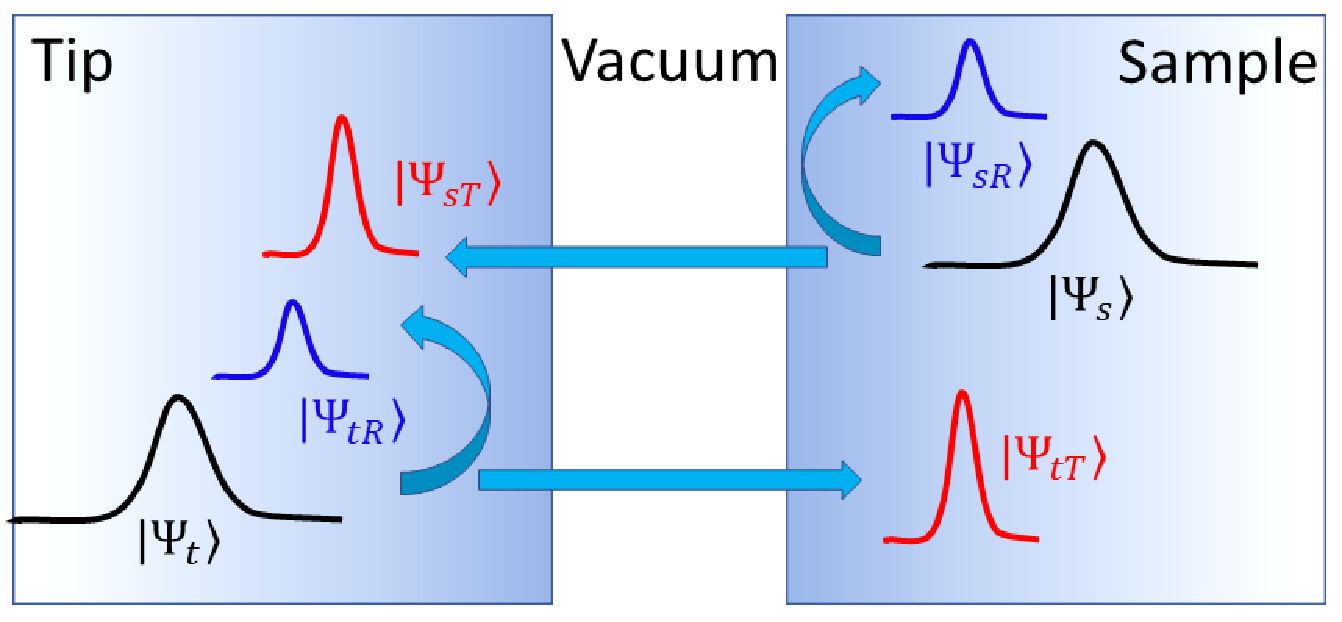}
\caption{\label{fig_S1} Wavefunctions in the tip and sample bifurcate into the reflected wavefunctions and tunneling wavefunctions separately at the final time. }
\end{figure*}

\section{\label{sec: 2}Derivation of the tunneling amplitude}

We define the field-free STM potential \DEL{as a rectangular barrier }with the Fermi energy $E_\mathrm{F, t}$ and $E_\mathrm{F, s}$ \DEL{and}\REV{as well as} work functions $W_\mathrm{t}$ and $W_\mathrm{s}$ for tip and sample~\cite{Yoshioka2016, Jelic2017}, respectively:
\begin{equation}\label{eq4} V_{0} = \begin{cases} -(E_\mathrm{F, t}+W_\mathrm{t}), &  x < 0 ,\\ V_\mathrm{imag}(x)-e(\varphi+U_\mathrm{s}) x/d, & 0 \leqslant x \leqslant d,\\ -(E_\mathrm{F, s}+W_\mathrm{s}+e\varphi+e U_\mathrm{s}), & x > d \end{cases}.
\end{equation}
\REV{where $e = -|e|$ is the charge of the electron, $V_\mathrm{imag}(x)$ is the image potential defined in ~\cite{Yoshioka2016}, $\varphi=(W_\mathrm{t}-W_\mathrm{s})/e$ is the contact potential difference, also known as Volta potential, and $U_\mathrm{s}$ is the static bias voltage applied to the STM. The tip boundary is set at $x=0$ and the sample boundary is at $x=d$.} Furthermore, we assume the tip and sample are ideal metals that can screen the laser field perfectly. The interaction $V_\mathrm{I}$ under the length gauge is given by

\begin{equation}\label{eq5} 
V_\mathrm{I}(t) = \begin{cases} 0, &  x < 0 ,\\ -e{\cal E}(t)x, & 0 \leqslant x \leqslant d,\\ -e{\cal E}(t)d, & x > d \end{cases}. 
\end{equation}
\DEL{where $e = -|e|$ is the charge of the electron. }\REV{The total potential with interaction in the Hamiltonian is $V(t)=V_{0}+V_\mathrm{I}(t)$. In the following, we will use $\chi_\mathrm{t}$, $\chi_\mathrm{gap}$, and $\chi_\mathrm{s}$ to select wavefunctions in the tip, gap, and sample, respectively. They will be 1 in the labeled regions and 0 outside these regions. The laser-induced tunneling is due to electrons driven across the gap by the laser field. Therefore, based on the Dyson equation, we can expand the evolution operator to the third order,} \DEL{To make derivations less complex, we further define two auxiliary potentials $V(t)=V_{0}+V_\mathrm{I}(t)$ and $V_\mathrm{s}(t)=V(t)+e{\cal E}(t)x$. Based on the expansion of the Dyson equation, we introduce the Volkov evolution operator $U_\mathrm{V}(\cdot,\cdot)$ into the time evolution operator,} 
\begin{eqnarray}\label{eq6}
U(t,-\infty)&=&U_{0}(t,-\infty)+\left(-\frac{i}{\hbar}\right)\int_{-\infty}^{t}U(t,t_{1})\chi_\mathrm{s}V_\mathrm{I}(t_{1})U_{0}(t_{1},-\infty)\;dt_{1}\,\nonumber\\
&&+\left(-\frac{i}{\hbar}\right)\int_{-\infty}^{t}U_\mathrm{Is}(t,t_{1})\chi_\mathrm{gap}V_\mathrm{I}(t_{1})U_{0}(t_{1},-\infty)\;dt_{1}\,\\
&&+\left(-\frac{i}{\hbar}\right)^{2}\int_{-\infty}^{t}\int_{-\infty}^{t_{2}}U(t,t_{2})[\chi_\mathrm{t}+\chi_\mathrm{s}][V(t_{2})-V_\mathrm{Is}(t_{2})]\nonumber\\
&& \ \ \ \times \, U_\mathrm{Is}(t_{2},t_{1})\chi_\mathrm{gap}V_\mathrm{I}(t_{1})U_{0}(t_{1},-\infty)\;dt_{1}dt_{2}\,.\nonumber
\end{eqnarray}
where $U_{0}(\cdot,\cdot)$ is the evolution operator with the interaction-free potential $V_{0}$. \REV{$V_\mathrm{Is}$ is the interaction with partially static potentials}

\begin{equation}\label{eq5a} 
V_\mathrm{Is}(t) = \begin{cases} 0, &  x < 0 ,\\ V_\mathrm{imag}(x)-e x [{\cal E}(t)+(\varphi+U_\mathrm{s})/d], & 0 \leqslant x \leqslant d,\\ -e[{\cal E}(t)d+\varphi+U_\mathrm{s}], & x > d \end{cases}. 
\end{equation}
\REV{and $U_\mathrm{Is}(\cdot,\cdot)$ is the evolution operator with $V_\mathrm{Is}$. The first term of Eq.~\ref{eq6} describes the evolution of the non-perturbed initial wave function. In contrast, the second term selects a perturbed region inside the sample, representing the initial wave function that has finished its tunneling before the laser comes. These two terms are unrelated to the laser-induced tunneling. They can be removed by the relation of Eq.~\ref{eq3}.}

When these operators work on the wave functions, we obtain the \REV{laser-induced} tunneling amplitude $M_{E}(t)$ as
\begin{eqnarray}\label{eq7}
M_{E}(t)&=&\langle\psi_{E}(t)\vert\chi_\mathrm{s}U(t,-\infty)\vert\Psi_{0}(-\infty)\rangle\\
&=&\langle\psi_{E}(t)\vert\chi_\mathrm{s}\vert\Psi_\mathrm{Is}(t)\rangle +\left(-\frac{i}{\hbar}\right)\int_{-\infty}^{t}\langle\psi(t_1)\vert[\chi_\mathrm{t}+\chi_\mathrm{s}][V(t_{1})-V_\mathrm{Is}(t_{1})]\vert\Psi_\mathrm{Is}(t_{1})\rangle\;dt_{1}\, ,\nonumber
\end{eqnarray}
where $\langle\psi(t_{1})\vert=\langle\psi_{E}(t)\vert\chi_\mathrm{s}U(t,t_{1})$ and $\vert\Psi_\mathrm{Is}(t)\rangle=(-\frac{i}{\hbar})\int_{-\infty}^{t}U_\mathrm{Is}(t,t_{1})\chi_\mathrm{gap}V_\mathrm{I}(t_{1})\Psi_{0}(t_{1})\;dt_{1}\,$. $\Psi_{0}$ is the initial wavefunction and $\psi_{E}$ is an eigenfunction of the sample with energy $E$.

Next, we use the following two equations,
\begin{eqnarray}\label{eq8}
\langle\psi(t)\vert(-i\hbar\frac{\partial}{\partial t}-H_{0})[\chi_\mathrm{t}+\chi_\mathrm{s}]=
\langle\psi(t)\vert V(t)[\chi_\mathrm{t}+\chi_\mathrm{s}]
\end{eqnarray}
and
\begin{eqnarray}\label{eq9}
[\chi_\mathrm{t}+\chi_\mathrm{s}](i\hbar\frac{\partial}{\partial t}-H_{0})\vert\Psi_\mathrm{Is}(t)\rangle&=&[\chi_\mathrm{t}+\chi_\mathrm{s}]V_\mathrm{Is}(t)\vert\Psi_\mathrm{Is}(t)\rangle
\end{eqnarray}
\REV{and apply them to Eq.~\ref{eq7}. In space representation, the integration of the second term of Eq.~\ref{eq7} is executed using partial integration in the following calculation,}
\begin{eqnarray}\label{eq7e}
&&\left(-\frac{i}{\hbar}\right)\int_{-\infty}^{t}\langle\psi(t_1)\vert[\chi_\mathrm{t}+\chi_\mathrm{s}][V(t_{1})-V_\mathrm{Is}(t_{1})]\vert\Psi_\mathrm{Is}(t_{1})\rangle\;dt_{1}\,\nonumber\\
&=&\int_{-\infty}^{t}\int_{-\infty}^{\infty}[\chi_\mathrm{t}+\chi_\mathrm{s}]\left[\Psi_\mathrm{Is}(x,t_1)\left(-\frac{\partial}{\partial t_1}-\frac{i\hbar}{2m}\frac{\partial^2}{\partial x^2}\right)\psi^*(x,t_1)-\psi^*(x,t_1)\left(\frac{\partial}{\partial t_1}-\frac{i\hbar}{2m}\frac{\partial^2}{\partial x^2}\right)\Psi_\mathrm{Is}(x,t_1)\right]\;dx\,\;dt_{1}\,\nonumber\\
&=&-\langle\psi(t)\vert[\chi_\mathrm{t}+\chi_\mathrm{s}]\vert\Psi_\mathrm{Is}(t)\rangle-\frac{i\hbar}{2m}\int_{-\infty}^{t}\int_{-\infty}^{\infty}[\chi_\mathrm{t}+\chi_\mathrm{s}] \frac{\partial}{\partial x}\left[\Psi_\mathrm{Is}(x,t_{1})\frac{\partial}{\partial x}\psi^{*}(x,t_{1})-\psi^{*}(x,t_{1})\frac{\partial}{\partial x}\Psi_\mathrm{Is}(x,t_{1})\right]\;dx\,\;dt_{1}\,\nonumber\\
&=&-\langle\psi_{E}(t)\vert\chi_\mathrm{s}[\chi_\mathrm{t}+\chi_\mathrm{s}]\vert\Psi_\mathrm{Is}(t)\rangle+\frac{i\hbar}{2m}\int_{-\infty}^{t} \left[\Psi_\mathrm{Is}(x,t_{1})\frac{\partial}{\partial x}\psi^{*}(x,t_{1})-\psi^{*}(x,t_{1})\frac{\partial}{\partial x}\Psi_\mathrm{Is}(x,t_{1})\right]\bigg\vert^{x=d}_{x=0}\;dt_{1}\,\\
&=&-\langle\psi_{E}(t)\vert\chi_\mathrm{s}\vert\Psi_\mathrm{Is}(t)\rangle+\frac{i\hbar}{2m}\int_{-\infty}^{t} \left[\Psi_\mathrm{Is}(x,t_{1})\frac{\partial}{\partial x}\psi^{*}(x,t_{1})-\psi^{*}(x,t_{1})\frac{\partial}{\partial x}\Psi_\mathrm{Is}(x,t_{1})\right]\bigg\vert^{x=d}_{x=0}\;dt_{1}\,.\nonumber
\end{eqnarray}
\REV{The first term of these results just cancels the first term in Eq.~\ref{eq7}, hence the simplified tunneling amplitude at an arbitrary time now reads}
\begin{eqnarray}\label{eq10}
M_{E}(t)&=&
\frac{i\hbar}{2m}\int_{-\infty}^{t} \left[\Psi_\mathrm{Is}(x,t_{1})\frac{\partial}{\partial x}\psi^{*}(x,t_{1})-\psi^{*}(x,t_{1})\frac{\partial}{\partial x}\Psi_\mathrm{Is}(x,t_{1})\right]\bigg\vert^{x=d}_{x=0}\;dt_{1}\,.
\end{eqnarray}
 \REV{The notation $[...]\big\vert^{x=d}_{x=0}$ at the end stands for the subtraction of the term inside the brackets evaluated at $x=0$ from the term evaluated at $x=d$.}
\DEL{The first term is unperturbed amplitude (static tunneling without laser field) and the second term is the laser-induced tunneling amplitude. In our case, which is symmetric,}\REV{For the tunneling from the sample, the amplitude is calculated in the same way but we need to use the opposite sign of the instantaneous laser field~\cite{Luo2021}.}

\begin{figure*}
\includegraphics[width=0.4\textwidth]{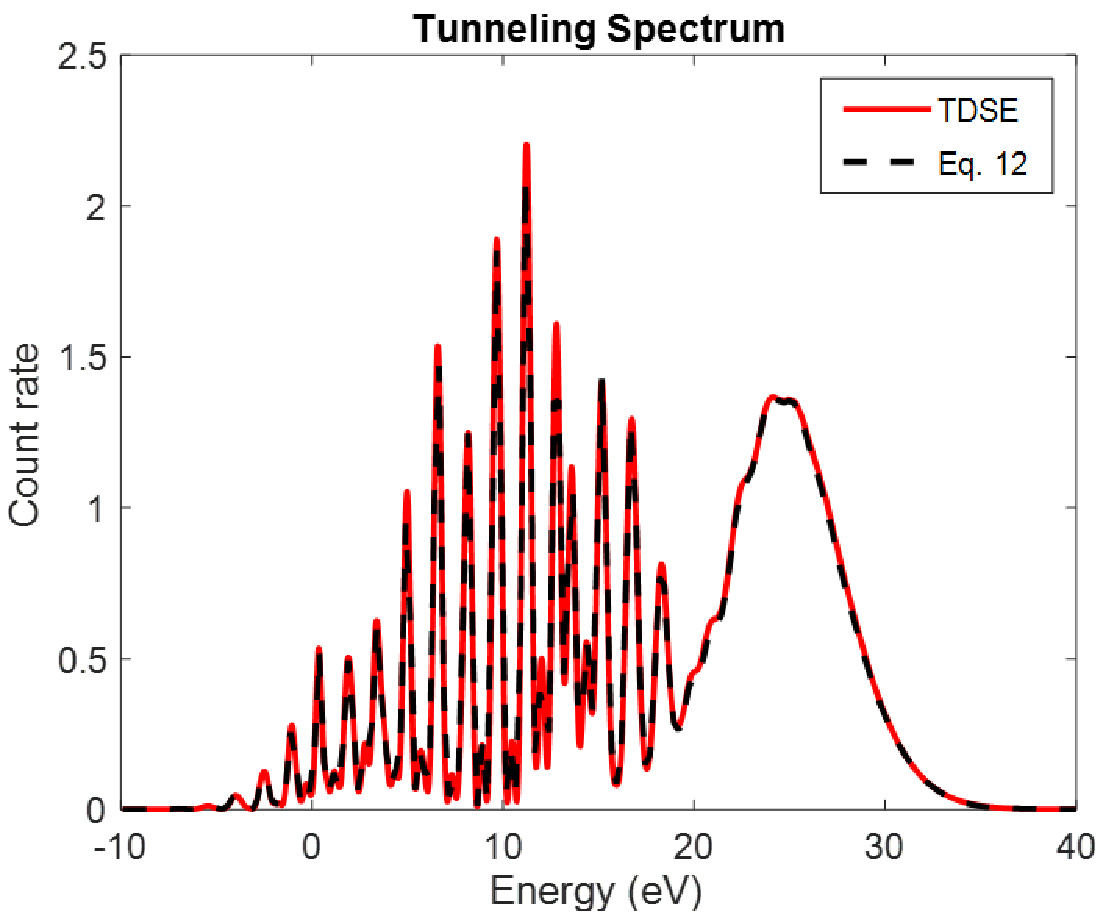}
\caption{\label{fig_S2} Comparison between TDSE simulation and the numerical calculation of Eq.~\ref{eq12}. We use a peak field of $F = 35\,\mathrm{V\,nm}^{-1}$ for a 6-fs laser pulse
at a central wavelength of 830\,nm. The width of the STM
junction is $d = 1\,$nm. \REV{Tip and sample are chosen to be identical (gold: $E_\mathrm{F,t} = E_\mathrm{F,s} = 5$\,eV, $E_0 = -5$\,eV, $W_\mathrm{t} = W_\mathrm{s} = 5$\,eV). The static bias voltage  $U_\mathrm{s}$ and the contact potential $\varphi$ are chosen to be zero.}  }
\end{figure*}

\DEL{and remove the unperturbed term from our equation.} \REV{For the accumulated tunneling current, the final measurement time can be set at infinity, when the interaction has been finished.} The \DEL{effective} \REV{laser-induced} tunneling magnitude is then given by
\begin{eqnarray}\label{eq12}
M_{E}&=&
\frac{i\hbar}{2m}\int_{-\infty}^{\infty} \left[\Psi_\mathrm{Is}(x,t_{1})\frac{\partial}{\partial x}\psi^{*}(x,t_{1})-\psi^{*}(x,t_{1})\frac{\partial}{\partial x}\Psi_\mathrm{Is}(x,t_{1})\right]\bigg\vert^{x=d}_{x=0}\;dt_{1}\,,
\end{eqnarray}
which has the same form as Bardeen's tunneling theory~\cite{Bardeen1961}. Figure~\ref{fig_S2} shows a comparison of Eq.~\ref{eq12} with the TDSE. The results are exactly identical.

By inserting the completeness operator into the current (Eq.~\ref{eq3}), \REV{we obtain the relation between tunneling amplitude Eq.~\ref{eq12} and the net current.}
\begin{eqnarray}\label{eq11}
{\cal J}&=&\langle \Psi_{tT} \vert\Psi_{tT}\rangle-\langle \Psi_{sT} \vert\Psi_{sT}\rangle\nonumber\\
&=&\sum_{E}[\langle \Psi_{tT} \vert\psi_{E}(\infty)\rangle\langle\psi_{E}(\infty)\vert\Psi_{tT}\rangle-\langle \Psi_{sT} \vert\psi_{E}(\infty)\rangle\langle\psi_{E}(\infty)\vert\Psi_{sT}\rangle]\\
&=&\sum_{E}(\vert M_{tE}\vert^2-\vert M_{sE}\vert^2).\nonumber
\end{eqnarray}

\section{\label{sec:3}Strong field approximation and saddle point equations}

By using the strong field approximation (SFA) introduced in the main text, the result of Eq.~\ref{eq12} is simplified to

\begin{eqnarray}\label{eq13}
M_{E,\mathrm{SFA}}&=&
\frac{i\hbar}{2m}\int_{-\infty}^{\infty} \left[\Psi_\mathrm{V}(x,t_{1})\frac{\partial}{\partial x}\psi^{*}_{E}(x,t_{1})-\psi^{*}_{E}(x,t_{1})\frac{\partial}{\partial x}\Psi_\mathrm{V}(x,t_{1})\right]_{x=d}\;dt_{1}\,.
\end{eqnarray}
\REV{where $\vert\Psi_\mathrm{V}(t)\rangle=(-\frac{i}{\hbar})\int_{-\infty}^{t}U_\mathrm{V}(t,t_{1})\chi_\mathrm{gap}V_\mathrm{I}(t_{1})\Psi_{0}(t_{1})\;dt_{1}\,$ is a superposition of the Volkov wave functions, and $U_\mathrm{V}(\cdot,\cdot)$ is the Volkov evolution operator.} To further simplify and compress this result, we consider an initial wave function at \REV{energy }$E_{0}$ \REV{ and zero bias voltage. The tip and sample are made from the same material (gold: $E_\mathrm{F,t} = E_\mathrm{F,s} = 5$\,eV, $W_\mathrm{t} = W_\mathrm{s} = -E_0 = 5$\,eV).}
\begin{equation}\label{eq14} \Psi_{0}(x) = \begin{cases} e^{i k_{1} x}+R e^{-i k_{1} x}, &  x < 0 ,\\ B_{1} e^{-\frac{1}{\hbar} \alpha x}+B_{2} e^{\frac{1}{\hbar} \alpha x}, & 0 \leqslant x \leqslant d,\\ T e^{i k_{3} x}, & x > d. \end{cases}
\end{equation}
where $\alpha=\sqrt{2m\vert E_{0}\vert}$. The Volkov propagator is defined as

\begin{eqnarray}\label{eq15}
\langle x_{2}\vert U_\mathrm{V}(t_{2},t_{1})\vert x_{1}\rangle=\sqrt{\frac{m}{2\pi i \hbar (t_2-t_1)}}e^{\frac{i}{\hbar}( p-eA(t_{2}))x_{2}}e^{-\frac{i}{\hbar}( p-eA(t_{1}))x_{1}}e^{-\frac{i}{\hbar}\frac{\int_{t_{1}}^{t_{2}}( p-eA(\tau))^2\;d\tau\,}{2m}},
\end{eqnarray}
where $ p\equiv\frac{\int_{t_{1}}^{t_{2}}{eA(\tau)}\;d\tau\,+m(x_{2}-x_{1})}{t_{2}-t_{1}}$. \REV{The static field in the gap can be included into the vector potential $A(t)=- \int^{t}_{-\infty\textbf{}} {\cal E}(\tau)\,d\tau$.}

We define two prefactors
\begin{eqnarray}\label{eq16}
\xi(t_{1})&\equiv&e{\cal E}(t_{1})(B_{1}\hbar^2\frac{-1+[1+\frac{i}{\hbar}(\tilde p-eA(t_{1})-i\alpha)d]e^{-\frac{i}{\hbar}(\tilde p-eA(t_1)-i\alpha)d}}{(\tilde p -eA(t_{1})-i\alpha)^2}\nonumber\\
&&+B_{2}\hbar^2\frac{-1+[1+\frac{i}{\hbar}(\tilde p-eA(t_{1})+i\alpha)d]e^{-\frac{i}{\hbar}(\tilde p-eA(t_1)+i\alpha)d}}{(\tilde p -eA(t_{1})+i\alpha)^2}),
\end{eqnarray}

\begin{eqnarray}\label{eq17}
\eta(t_{2})&\equiv&(\tilde p -eA(t_{2})+\sqrt{2m(E-U_\mathrm{s})})e^{-\frac{i}{\hbar}\sqrt{2m(E-U_\mathrm{s})}d},
\end{eqnarray}
 the effective canonical momentum $\tilde p=\frac{\int_{t_{1}}^{t_{2}}{eA(\tau)}\;d\tau\,+md}{t_{2}-t_{1}}$, and a phase factor based on the semi-classical action

\begin{eqnarray}\label{eq18}
S(t_{2},t_{1})&\equiv&E t_{2}+\frac{{\tilde p}^2}{2m}(t_{2}-t_{1})-\int_{t_{1}}^{t{2}}\frac{e^2A^2(\tau)}{2m} \;d\tau\,+\vert E_{0}\vert t_{1}.
\end{eqnarray}

With all this in hand, we derive the laser-driven tunneling amplitude $M_{E,\mathrm{SFA}}$ for an initial wavefunction at energy $E_{0}$ as

\begin{eqnarray}\label{eq19}
M_{E,\mathrm{SFA}}=\int_{-\infty}^{\infty}\int_{-\infty}^{t_{2}}\sqrt{\frac{i}{8\pi m\hbar^3(t_{2}-t_{1})}}
\eta(t_{2})\xi(t_{1})e^{\frac{i}{\hbar}S(t_{2},t_{1})}\;dt_{1}dt_{2}\,.
\end{eqnarray}

This temporal integration contains a rapidly oscillating function $e^{\frac{i}{\hbar}S(t_{2},t_{1})}$ so that it can be analyzed by the saddle point technique. To find the zero points of the phase derivatives $\frac{\partial S(t_2,t_1)}{\partial t_{1}}\big\vert_{t_{1}=t_{1s}}=\frac{\partial S(t_2,t_1)}{\partial t_{2}}\big\vert_{t_{2}=t_{2s}}=0$, we can retrieve the saddle point equations

\begin{eqnarray}
\frac{\left[\tilde p_\mathrm{s}-eA(t_\mathrm{1s})\right]^2}{2m}=-\left\vert E_0\right\vert,
\\
\int_{t_\mathrm{1s}}^{t_\mathrm{2s}}\frac{\left[\tilde p_\mathrm{s}-eA(\tau)\right]}{m}\;d\tau\,=d,
\\
\frac{\left[\tilde p_\mathrm{s}-eA(t_\mathrm{2s})\right]^2}{2m}=E.
\end{eqnarray}
The subscript (s) indicates that the value is a saddle point.

\section{\label{sec:4}Classical trajectory and cutoff energy}

The saddle point equations can be used to formulate semi-classical trajectories of the tunneling electrons and their transport to the sample region. In Newtonian mechanics, the displacement of a classical point-like electron in the laser field is
\begin{eqnarray}\label{eq23}
{\cal D}(t)=\int_{t_\mathrm{1s}}^{t}\frac{\left[\tilde p_\mathrm{s}-eA(\tau)\right]}{m}\;d\tau\,.
\end{eqnarray}
In our case, the negative initial energy in the saddle point equation causes all the intermediate values to be complex. Therefore, this displacement must be integrated along a contour on the complex plane, in analogy to saddle point equations for HHG, for instance~\cite{Smirnova2014}.

\begin{figure*}
\includegraphics[width=0.5\textwidth]{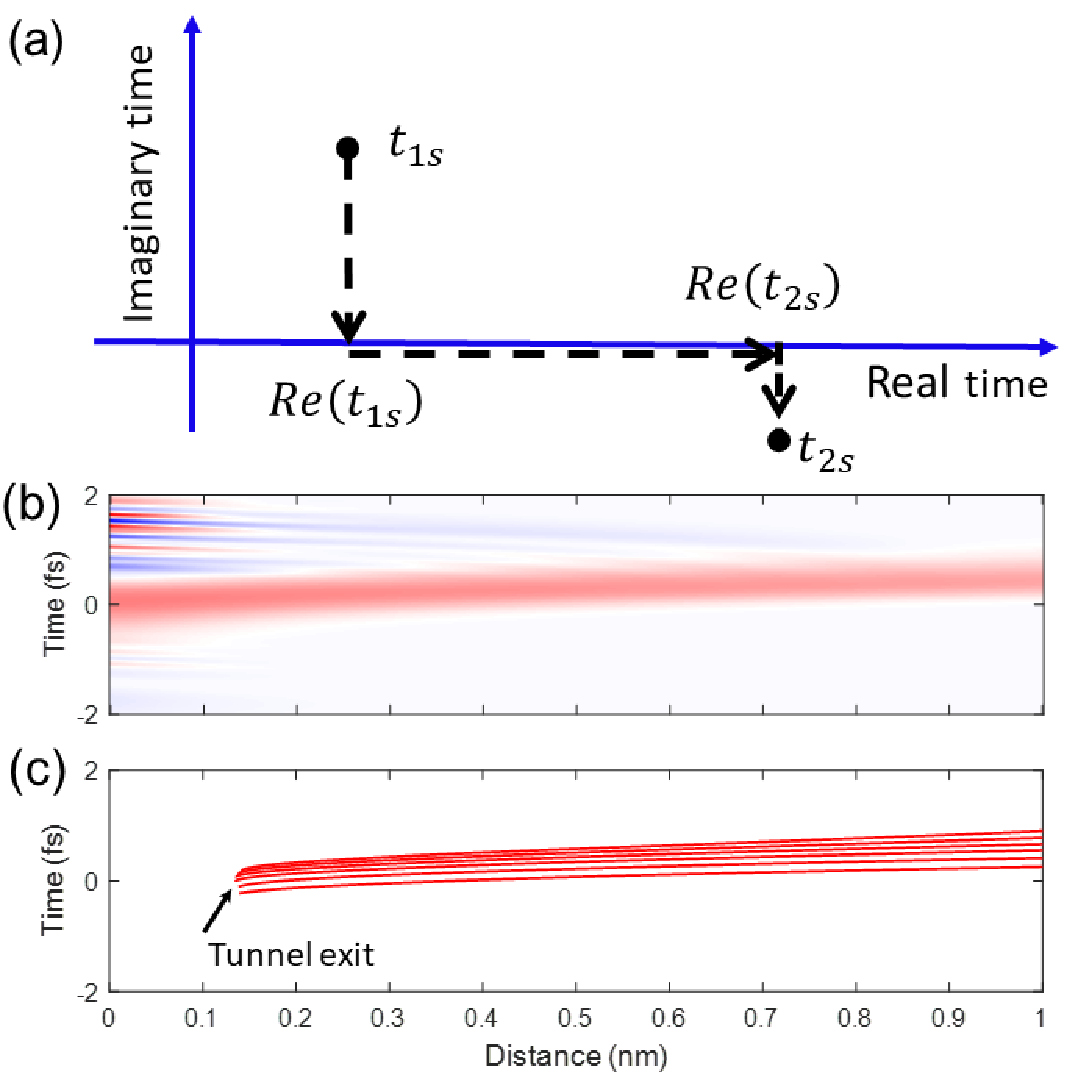}
\caption{\label{fig_S3} (a) Integation contour on the complex plane for the semi-classical trajectories. (b) and (c) Comparison between the TDSE trajectory and the superposition of several possible classical trajectories. In these simulations, the junction width is $d = 1\,$nm, the pulse duration is 3\,fs (one cycle), the field strength is $35\,\mathrm{V\,nm}^{-1}$, and the central wavelength is 830\,nm.}
\end{figure*}

Figure~\ref{fig_S3}(a) shows our integration contour. It contains three physical processes. The first contour from the initial time $t_\mathrm{1s}$ to its real part is tunneling, a classically forbidden process. The second contour along the real axis is the classical trajectory inside the junction. The third contour which is from the real axis to the final $t_\mathrm{2s}$ corresponds to the attenuation during transmission.

Figure~\ref{fig_S3}(b) and (c) show the current burst based on the TDSE and the superposition of possible classical trajectories, respectively. The void space near $x = 0$ in Figure~\ref{fig_S3}(c) is the region where the tunneling process occurs. The trajectories emerge at the tunnel exit.

The cutoff energy is the maximum kinetic energy that a classical electron can gain from the laser field in the process. According to the TDSE simulations shown in Figure~\ref{fig_S3}(b), the central electron burst is around the crest of the laser waveform when the instantaneous field is strongest ($t=0$). Therefore, we can solve saddle point equations near the crest by using trigonometric functions with central frequency $\omega$ of the incident laser. The electric field is described as

\begin{eqnarray}\label{eq24}
{\cal E}(t)=-F_{0}-F \cos (\omega t),
\end{eqnarray}
where $F_{0}$ is the static field strength from the bias voltage \REV{and the contact (Volta) potential}, and $F$ is the field strength of the incident laser. The vector potential is given by

\begin{eqnarray}\label{eq25}
A(t)=F_{0}t+\frac{F}{\omega}\sin (\omega t).
\end{eqnarray}

For a lower energetic state, the electron will take more travel time $\tau=t_\mathrm{2s}-t_\mathrm{1s}$ to be transported over the junction. In this case, the imaginary part of $\tau$ can be approximated with the well-known Keldysh parameter~\cite{Keldysh1965,Pedatzur2015},

\begin{eqnarray}\label{eq26}
i\omega \Im(\tau)=-\frac{i\omega\sqrt{2m|E_{0}|}}{|e|(F_{0}+F)}.
\end{eqnarray}

For a higher energetic state, the freed electron can take a short time to travel across the junction, which causes $\tau$ to approach zero. We can use a Taylor expansion to express the duration for this case,

\begin{eqnarray}\label{eq27}
\tau=\frac{2md}{\sqrt{2mE}+i\sqrt{2m|E_{0}|}}.
\end{eqnarray}

\begin{figure*}
\includegraphics[width=0.4\textwidth]{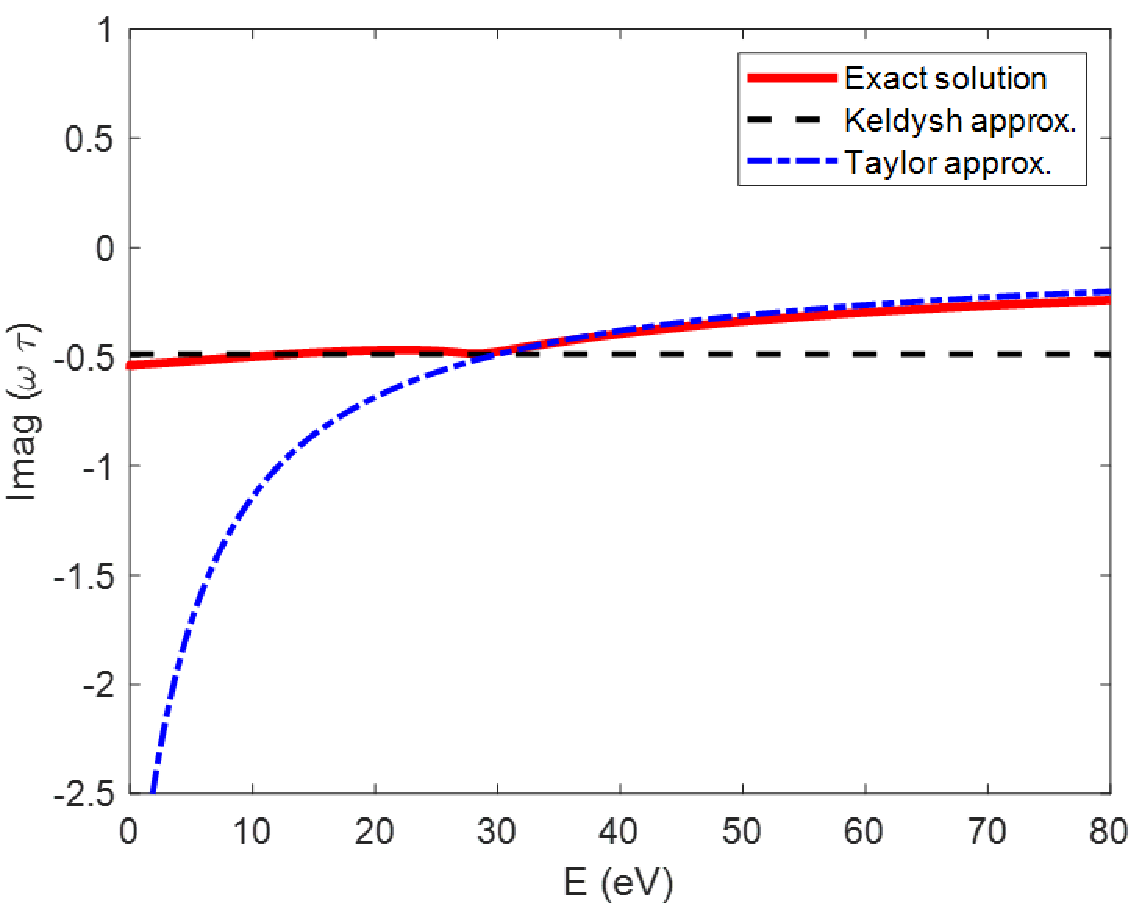}
\caption{\label{fig_S4} Comparison between two approximated solutions for the imaginary part of the travel time $\tau$ with the exact numerical results (red).}
\end{figure*}

Figure~\ref{fig_S4} shows these two approximated solutions for the imaginary part of $\tau$ as a function of final energy $E$ in one plot, together with the exact numerical solution. Each of the two solutions agrees well with the exact solution in two separate spectral regions. We note that Eq.~\ref{eq26} has a constant imaginary value and Eq.~\ref{eq27} exhibits a changing imaginary value. In the latter case, the increase of the imaginary part of the travel time leads to a strong attenuation of the tunneling amplitude. Therefore, we define the intersection of these two approximated solutions as the cutoff energy. Equating the imaginary part of Eq.~\ref{eq26} and Eq.~\ref{eq27}, we obtain the cutoff energy
\begin{eqnarray}\label{eq28}
E_\mathrm{cutoff}=|e|(F_{0}+F)d-|E_{0}|.
\end{eqnarray}

\section{\label{sec:5}Carrier-envelope phase modulation of the tunneling current}
For a few-cycle laser pulse, the carrier-envelope phase (CEP) modulation can significantly break the system's symmetry, rectifying the current flow in the junction. The CEP control in the ultrafast STM can be implemented in various conditions such as different junction widths and laser pulse durations. Here, we use the TDSE to test the efficiencies in these conditions. As we have shown in the main text, the CEP effect is only present for sufficiently short laser pulses and is  averaged out when the pulse duration becomes large.

\begin{figure*}
\includegraphics[width=0.7\textwidth]{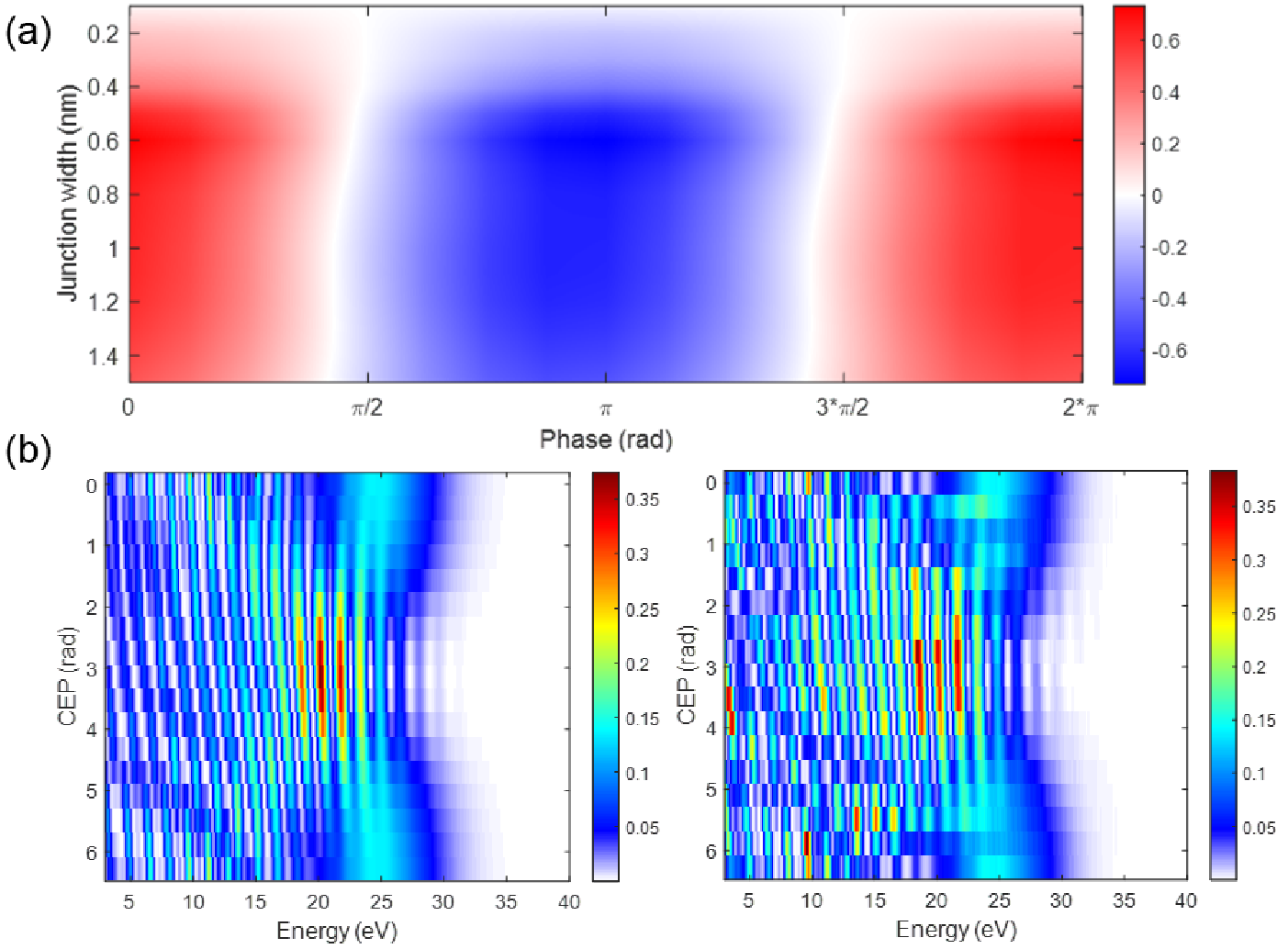}
\caption{\label{fig_S5} (a) The laser-induced current as a function of CEP and junction width. For a width range from 0.2 nm to 1.5 nm, CEP can efficiently reverse the tunneling currents. In this TDSE simulation, we use 3-fs (one cycle) of field strength $F = 17\,\mathrm{V\,nm}^{-1}$ laser pulse at 830 nm central wavelength. (b) Energy spectra which are simulated with TDSE (left) and SFA (right) for a 6-fs pulse. The cutoff energies sinusoidally change with respect to the CEP. In these simulations, the field strength is set as $ 35\,\mathrm{V\, nm}^{-1}$ and the junction width $d$ is 1 nm.  }
\end{figure*}

The width of the STM junction is another factor that can affect the CEP modulation. In Figure.~\ref{fig_S5}(a), we show that the CEP modulation is quite stable when we retract the tip away from the sample from 0.2 nm to 1.5 nm. This range can basically cover most STM experiments. Even smaller junctions could cause the tip damage. It is also evident that the CEP effect is weak for $d < 0.5$\, nm, hence larger junction widths are preferred for inducing single attosecond bursts of tunneling electrons.

\begin{figure*}
\includegraphics[width=0.8\textwidth]{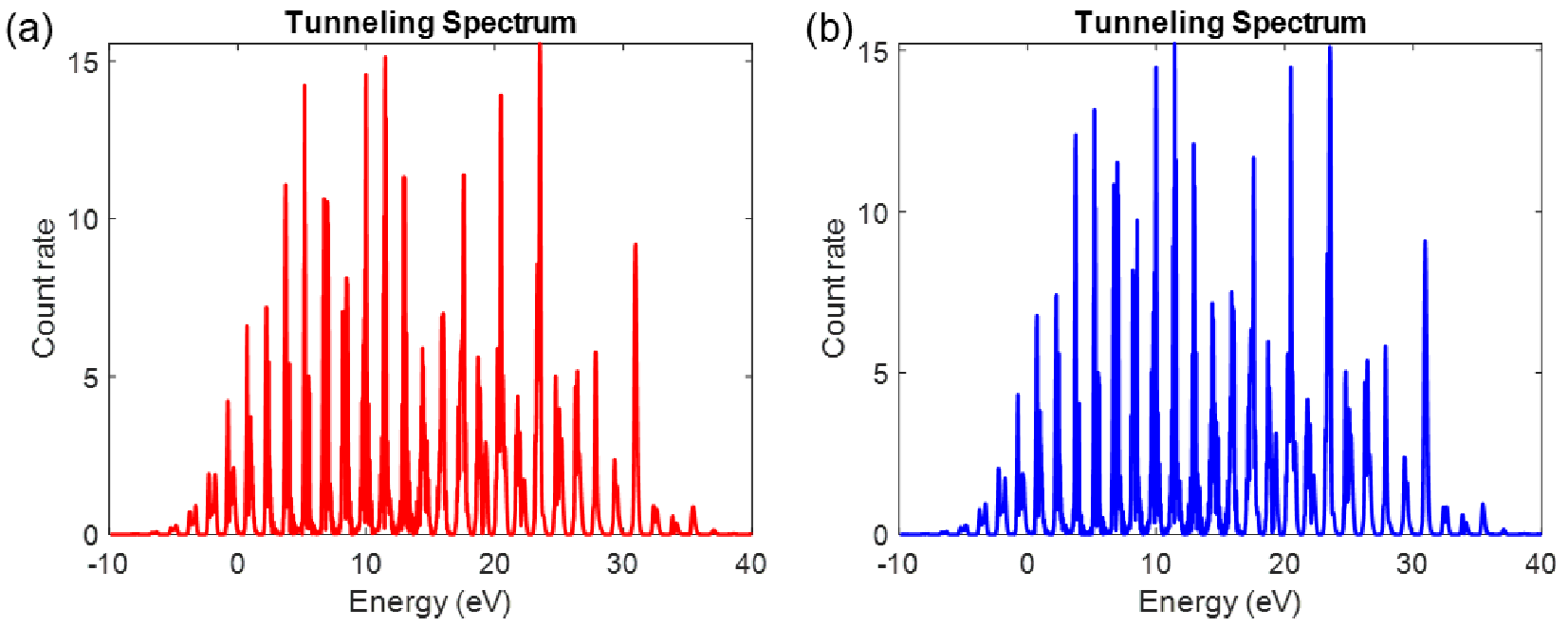}
\caption{\label{fig_S5a} Comparison of the tunneling spectra from the tip to the sample (a) and from the sample to the tip (b). In these TDSE simulations, we use a 50-fs (18 cycles) laser pulse of field strength $F = 35\,\mathrm{V\,nm}^{-1}$ at 830 nm central wavelength and a junction width $d$ of 1\,nm.  }
\end{figure*}

The CEP modulation can be also interpreted from the energy spectrum based on the SFA. Figure.~\ref{fig_S5}(b) shows a comparison between the TDSE simulation and SFA calculation for a 6-fs laser pulse. The high-energy parts of the spectra are modulated strongly with respect to the CEP. The cutoff energy obtained from the SFA shows linear dependence with field strength $F$. For short laser pulses, the shift in the CEP changes the effective field strength sinusoidally, leading to these periodical cutoff modulations.  Also, interference fringes appear and disappear with CEP at 25\,eV near the cutoff, which is a signature of temporal double-slit and single-slit interference, respectively~\cite{Kruger2011,Ott2013}. \REV{Furthermore, as what we can see from Fig.~4(b) in the main text, the tunneling current can vanish for long pulses. Figure~\ref{fig_S5a} shows the tunneling spectra from the two sides are almost indistinguishable when the pulse duration is 50\,fs (18 cycles).}

\section{\label{sec:6}Numerical calculations of the time-dependent Schr\"odinger equation (TDSE)}

\REV{In the main text in Fig.~2 and 3, we compare the SFA results with numerical solutions of the TDSE in exactly the same geometry and physical situation. The one-dimensional TDSE is solved using the Crank-Nicolson method, which has been widely used for strong-field physics simulations at nanotips~\cite{Hommelhoff2006, Yalunin2011, Kruger2011, Dienstbier2023}. The Hamiltonian in the TDSE is 
\begin{eqnarray}\label{eq29}
H=-\frac{\hbar^{2}}{2m}\frac{\partial^2}{\partial x^2}+V_{0}(x)+V_\mathrm{I}(x),
\end{eqnarray}
where field-free potential $V_{0}(x)$ and interaction $V_\mathrm{I}(x)$ are defined in Eq.~\ref{eq4} and Eq.~\ref{eq5}. In our simulations, we assume that the tip and the sample are made from gold, with a Fermi energy of 5\,$\mathrm{eV}$ and a work function of 5\,$\mathrm{eV}$. The spatial grid and the temporal step size are determined by uncertainty relations
\begin{eqnarray}\label{eq30}
\Delta p \cdot \Delta x \geqslant \frac{\hbar}{2},
\end{eqnarray}
\begin{eqnarray}\label{eq31}
\Delta E \cdot \Delta t \sim \hbar.
\end{eqnarray}
In our simulations, the maximum expected cutoff energy is 30 $\mathrm{eV}$ ($F=35\,\mathrm{V\,nm}^{-1}$, $d=1\,\mathrm{nm}$). Therefore, we set a maximum energy bandwidth $\Delta E=300\,\mathrm{eV}\gg 30\,\mathrm{eV}$ and apply $\Delta x \rightarrow dx$ and $\Delta t \rightarrow dt$ (here: $\Delta x = 0.01\, \mathrm{nm} $ and $\Delta t = 2.2\, \mathrm{as}$). The incident few-cycle laser electric field has a Gaussian envelope function and its vector potential is given by
\begin{eqnarray}\label{eq32}
A(t)=\frac{F}{\omega}e^{-4 \ln 2(t / \tau)^2}\sin{(\omega t+\phi)},
\end{eqnarray}
where $F$ denotes the peak field strength and $\omega$ is the angular frequency. $\mathrm{\tau}$ and $\phi$ are the pulse duration (FWHM) and the carrier-envelope phase (CEP), respectively. In our simulations, the laser parameters are $\omega=2.27\times10^{15} 
\,\mathrm{s^{-1}}$ ($\lambda = 830\,\mathrm{nm}$) and $\tau=6 \,\mathrm{fs}$. The calculation time is four times as long as the pulse duration ($-12\,\mathrm{fs} \leqslant t \leqslant 12\,\mathrm{fs}$). }

\REV{In order to avoid undesired reflections of the wave function at the end of the spatial grid, we terminate the grid at positions far away from the tunneling wave functions. A classical point-like electron with a kinetic energy of 30\,$\mathrm{eV}$ (maximum cutoff energy) can only travel around 40\,$\mathrm{nm}$ within 12\,$\mathrm{fs}$. The termination positions are set at 300\,$\mathrm{nm}$, almost 8 times as far away as the fastest electron. The tunneling wave functions are obtained by subtracting the wave function in the sample domain at the initial time from the wave function at the final time. Since the eigenfunctions inside the sample have analytical solutions, we project our simulated tunneling waves directly onto these eigenstates. }


%